\documentclass[reprint, amsmath, amssymb, aps]{revtex4-2}
\usepackage{graphicx}
\usepackage{amssymb}
\usepackage{dcolumn}
\usepackage{bm}
\usepackage{dcolumn}

\def\Ra{R_{\mathrm{act}}}
\def\Ri{R_{\mathrm{inh}}}

\def\bea{\begin{eqnarray}}
\def\eea{\end{eqnarray}}
\def \be{\begin{equation}}
\def \ee{\end{equation}}

\begin{document}
\title{Trait-space patterning and the role of feedback in antigen-immunity coevolution}
\author{Hongda Jiang}
\author{Shenshen Wang}
\email{shenshen@physics.ucla.edu}
\affiliation{Department of Physics and Astronomy, University of California Los Angeles, Los Angeles, CA 90095, USA}

\begin{abstract}
Coevolutionary arms races form between interacting populations that constitute each other's environment and respond to mutual changes. This inherently far-from-equilibrium process finds striking manifestations in the adaptive immune system, where highly variable antigens and a finite repertoire of immune receptors coevolve on comparable timescales. This unique challenge to the immune system motivates general questions: How do ecological and evolutionary processes interplay to shape diversity? What determine the endurance and fate of coevolution? Here, we take the perspective of responsive environments and develop a phenotypic model of coevolution between receptors and antigens that both exhibit cross-reactivity (one-to-many responses). The theory predicts that the extent of asymmetry in cross-reactivity
is a key determinant of repertoire composition: small asymmetry supports persistent large diversity, whereas strong asymmetry yields long-lived transients of quasispecies in both populations. The latter represents a new type of Turing mechanism.
More surprisingly, patterning in the trait space feeds back on population dynamics: spatial resonance between the Turing modes breaks the dynamic balance, leading to antigen extinction or unrestrained growth.
Model predictions can be tested via combined genomic and phenotypic measurements.
Our work identifies cross-reactivity as an important regulator of diversity and coevolutionary outcome, and reveals the remarkable effect of ecological feedback in pattern-forming systems, which drives evolution toward non-steady states different than the Red Queen persistent cycles.
\end{abstract}

\maketitle




\section{Introduction}

Highly variable antigenic challengers, such as fast evolving viruses and cancer cells, are rapid in replication and abound with genetic or phenotypic innovations~\cite{duffy:08, greaves:12}, thus managing to evade immune recognition. On the reciprocal side, the host's immune system adjusts on the fly the clonal composition of its finite receptor repertoire to recognize the altered versions of the antigen. As a result, coevolutionary arms races between antigen and immunity may endure through an individual's lifetime~\cite{liao:13}.

In this Red Queen~\cite{valen:73} scenario, antigen and receptor populations constitute each other's responsive environment and are mutually driven out of equilibrium: specific immune receptors prey on matching antigens and hence alter both the composition and overall abundance of antigens, which in turn modifies selective pressures on distinct receptors thus causing re-organization of the repertoire, and vice versa.
Consequently, neither of the populations has enough time to equilibrate and yet they mutually engage in a dynamic balance. In this sense, the Red Queen state represents a nonequilibrium steady state~\cite{lax:60, qian:06}. Then the question is whether alternative evolutionary outcomes characteristic of non-steady states would occupy a larger volume of the state space of coevolving systems than does the Red Queen state.

Recent progress has been made toward understanding various aspects of coevolutionary dynamics in antigen-immunity systems~\cite{luo:15, armita:16, cobey:15, zanini:15, neher:16, luksza:14, luksza:17, weitz:05, bradde:17}, ranging from antibody evolution against HIV and influenza viruses to evolution of tumors and bacterial phage under host immunity. Yet we are still short of insights regarding key fundamental questions: How do receptor repertoire and antigen ensemble mutually organize, when ecological and evolutionary dynamics occur on comparable timescales? What govern the persistence and outcome of mutual adaptation?
In existing generic models where both ecological and evolutionary processes are considered, a separation of timescales is often assumed so that the fast dynamics is slaved to the slow one (reviewed in~\cite{dieckmann:04}). In cases where timescales are not treated as separated~\cite{fussmann:00, yoshida:07, jones:07, shih:14, vetsigian:17, wienand:18, halatek:18, kotil:18}, the feedback between changes in diversity and population dynamics tends to be ignored.
The goal of this paper is to consider inseparable timescales and at the same time account for feedback effect in order to address the questions raised above.

Specifically, we develop a phenotypic model, based on predator-prey interactions between coevolving immune receptors and antigens, that combines evolutionary diversification and population dynamics.
By formulating an ecological model in a trait space, we describe coevolutionary changes in the distribution of trait values and trait-dependent predation in the same framework. Importantly, this allows us to study the stability of speciation (pattern formation in the trait space) and its impact on the persistence of coevolution.
Our model abstracts the key features of adaptive immunity:
antigens and receptors move (due to trait-altering mutations) and behave like activators and inhibitors that react through predation; both antigens and receptors are cross-reactive --- one receptor recognizes many distinct antigens and one antigen is recognized by multiple receptors --- this flexibility in recognition stems from structural conservation of part of the receptor/antigen binding surface~\cite{birnbaum:14} and provides an enormous functional degeneracy~\cite{wooldridge:12} among distinct immune repertoires; there is no preexisting fitness landscape for either population so that selection pressures are owing purely to predation.

The theory predicts, counterintuitively, that simultaneous patterning in coevolving populations can emerge solely from asymmetric range of activation and inhibition in predator-prey dynamics, without a need for competitive interaction within either population~\cite{scheffer:06, pigolotti:07, doebeli:10, rogers:12} or severely large differences in their rate of evolution~\cite{turing:52} (aka mobility in their common phenotypic space), thus representing a new Turing mechanism.
This surprising result can be understood from an intuitive picture:
colocalized clusters of antigens and receptors form in the trait space when the ``inhibition radii" of adjacent receptor clusters overlap so that inhibition of antigen is strongest in between them; whereas alternate clusters emerge if the ``activation radii" of neighboring antigen clusters intersect, because then activation of receptor is most intense in the midway.
We show for the first time that, by varying the asymmetry in cross-reactivity alone, transitions between qualitatively distinct regimes of eco-evolutionary dynamics observed in nature would follow, including persistent coexistence, antigen elimination and unrestrained growth.
A new interesting outcome of our analysis is, spatial resonance of Turing modes can drive the arms races off balance: given sufficient asymmetry, spontaneous oscillations in Turing patterns precede antigen extinction, whereas uncontrolled antigen growth follows the formation of alternate quasispecies as ineffective receptors exhaust the limited immune resources; these measurable features may serve as the precusors of the off-balance fates.


Many theoretical studies have considered adaptation to time-varying environments with prescribed environmental statistics~\cite{meyers:02, kussell:05, mustonen:08, mustonen:09, ivana:15, xue:16, xue:19}. This work makes a step toward a theory of coevolution from the perspective of responsively changing environments (mutual niche construction~\cite{smee:96} in ecological terms), highlighting the role of feedback in driving evolution toward novel organization regimes and non-steady states.
As new genomic and phenotypic methods are developed to better characterize antigenic~\cite{bloom:06, gong:13, wu:15}
and immunogenic~\cite{klein:13, diskin:13, boyer:16, adams:16} landscapes as well as bidirectional cross-reactivity~\cite{west:13}, the predictions for repertoire composition and coevolutionary outcome derived from this study can be compared with high-throughput profiling of coevolving immune repertoire and antigen ensemble in humans~\cite{gao:14, moore:15, freund:17}.

\section{Model}
A finite repertoire of immune receptors that collectively cover the antigenic space while leaving self types intact is conceivable, if the distribution of potential threats is predetermined~\cite{deboer:94, deboer:01, mayer:15, wang:17}.
Given a fixed distribution of pathogenic challenges, competitive exclusion is shown to drive clustering of cross-reactive receptors~\cite{mayer:15}.
In coevolution, however, antigen distribution is no longer preset but responds to reorganization of receptors. In addition,
cross-reactivity is bidirectional: not only can a receptor be activated by a range of distinct antigens, but an antigen can be removed by a variety of receptors.
Then, can predation lead to simultaneous clustering of antigens and receptors in their common trait space? If so, are such patterns stable? Would the concurring patterns interact to affect population dynamics?

\begin{figure}[t]
  \includegraphics[width=1\columnwidth]{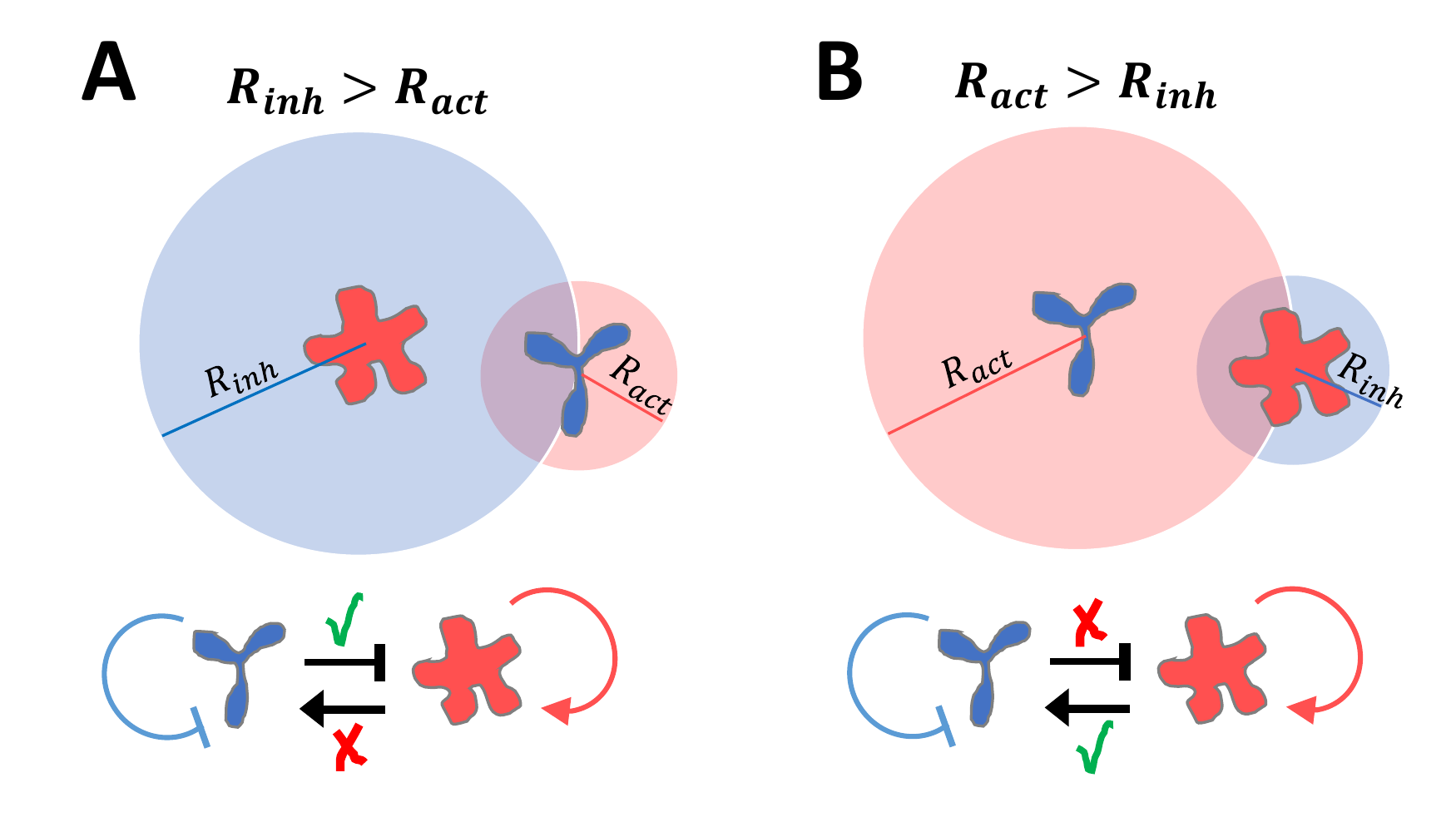}
  \caption{Schematic of antigen-receptor interaction with asymmetric range of inhibition and activation in the phenotypic space. (A) $\Ri>\Ra$: the receptor (blue Y-shape) is not activated by the antigen (red flower-shape) but nevertheless inhibits it. (B) $\Ra>\Ri$: the antigen activates the receptor but is not subject to its inhibition. Lower row: in addition to predation (black arrows; blunt for inhibition, acute for activation), antigens self replicate (red acute arrow) whereas receptor-carrying cells spontaneously decay (blue blunt arrow) in the absence of stimulation.
  }
\label{asymmetry}
\end{figure}
To answer these questions, we consider a dynamical system of activators and inhibitors representing antigens and receptors, which diffuse in a shared phenotypic space and react through predator-prey interactions. Population densities of antigens $A(\vec{x},t)$ and receptors $B(\vec{x},t)$ evolve according to
\begin{eqnarray}
\partial_tA(\vec{x},t)&=&D_1\nabla^2A(\vec{x},t)+\lambda_1A(\vec{x},t)\nonumber\\
&-&\alpha_1A(\vec{x},t)\int S_1(|\vec{x}-\vec{y}|;\Ri)B(\vec{y},t)\mathrm{d}\vec{y},\nonumber\\
\partial_tB(\vec{x},t)&=&D_2\nabla^2B(\vec{x},t)-\lambda_2B(\vec{x},t)+B_{\mathrm{in}}\nonumber\\
&+&\alpha_2B(\vec{x},t)\int S_2(|\vec{x}-\vec{y}|;\Ra)A(\vec{y},t)\mathrm{d}\vec{y}.
\end{eqnarray}
Here, $D_1$ and $D_2$ denote isotropic diffusion constants of antigens and receptors, respectively, that mimic the rates of trait-altering mutations. Other forms of jump kernels do not change qualitative results [SI]. Antigens self replicate at rate $\lambda_1$ whereas receptors spontaneously decay at rate $\lambda_2$. Receptors inhibit antigens with an intrinsic rate $\alpha_1$ and grow at rate $\alpha_2$ upon activation; $\Ri$ denotes the range of receptors that can inhibit a given antigen, while $\Ra$ represents the range of antigens by which a receptor can be activated (Fig.~\ref{asymmetry}). In real systems, there is likely a distribution of reaction range; we assume a single value to simplify analysis. The term $B_{\mathrm{in}}$ corresponds to a small influx of lymphocytes that constantly output from the bone marrow and supply nascent receptors; without stimulation, receptors are uniformly distributed at a resting concentration given by $B_{\mathrm{in}}/\lambda_2$.
We choose the lifetime of receptors, $\lambda_2^{-1}$, as the time unit and the linear dimension $L$ of the phenotypic space as the length unit.
To account for the discreteness of replicating entities and hence avoid unrealistic revival from vanishingly small population densities, we impose an extinction threshold; antigen or receptor types whose population falls below this threshold are considered extinct and removed from the system.


In the spirit of Perelson and Oster~\cite{perelson:79}, we think of receptors and antigens as points in a high-dimensional phenotypic shape space, whose coordinates are associated with physical and biochemical properties that affect binding affinity. We assume that the strength of cross-reactive interaction only depends on the relative location, $\vec{r}=\vec{x}-\vec{y}$, of receptor and antigen in this space, as characterized by the non-local interaction kernels $S_1(|\vec{r}|; \Ri)$ and $S_2(|\vec{r}|; \Ra)$. Close proximity in shape space indicates good match between the binding pair leading to strong interaction, whereas large separation translates into weak affinity and poor recognition.
Importantly, cross-reactivity is not necessarily symmetric as typically assumed; difference in biophysical conditions among other factors may well render disparate criteria for antigen removal and receptor activation~\cite{tarr:11, sather:14}, i.e., $\Ri\neq\Ra$. For instance, removing an antigen may only require modest on rate (wide reaction range, large $\Ri$) of multiple receptors that together coat its surface, whereas activating an immune cell expressing a unique type of receptors can demand lasting antigen stimulation hence small off rate (close match of shape, small $\Ra$), or vice versa depending on conditions.
How this asymmetry impacts coevolution is our focus.


\begin{figure}[t]
  \includegraphics[width=0.85\columnwidth]{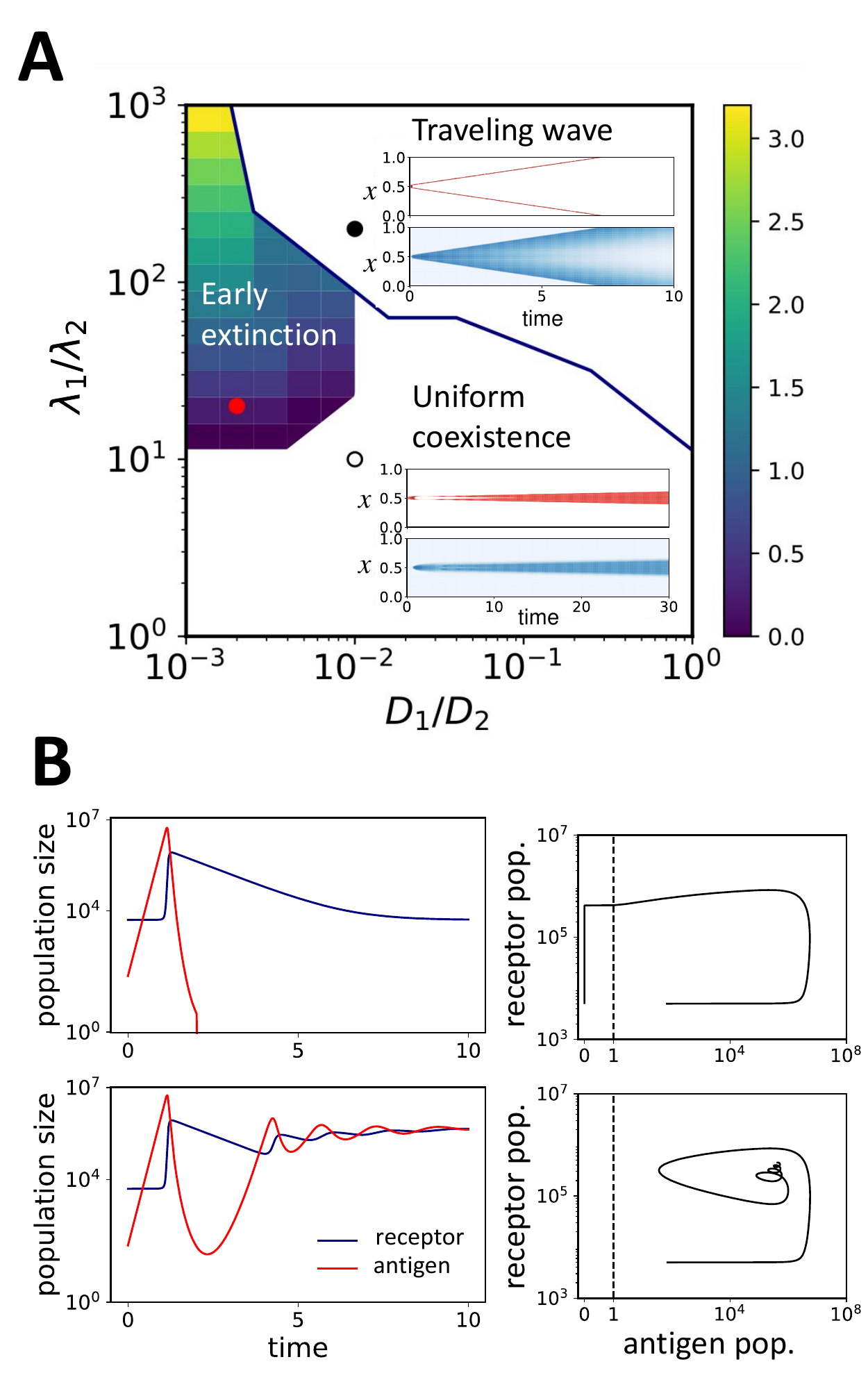}
  \caption{Phases in a 1D reaction-diffusion system under local predator-prey interactions. (A) Phase diagram on the plane spanned by the ratio between diffusion constants $D_1/D_2$ and that between birth and death rates $\lambda_1/\lambda_2$ of antigens (activators) and receptors (inhibitors).
  Dynamics start from a local dose of antigens and uniform receptors. The early extinction phase is color coded for the logarithm of the inverse time to antigen extinction. The persistence phase (blank) divides into a propagating wave state (upper) and a uniform coexistence state (lower).
  Insets show typical kymographs in each subphase, red for antigen and blue for receptor; the upper pair corresponds to the filled circle at $\lambda_1/\lambda_2=200$, $D_1/D_2=10^{-2}$, and the lower one corresponds to the open circle at $\lambda_1/\lambda_2=10$, $D_1/D_2=10^{-2}$.
  (B) Representative abundance trajectories. Top: $\lambda_1/\lambda_2=20$, $D_1/D_2=10^{-3}$ (red dot in panel A); bottom: $\lambda_1/\lambda_2=10$, $D_1/D_2=10^{-2}$. Corresponding phase plots are shown on the right; vertical dashed lines indicate the extinction threshold. $B_{\mathrm{in}}=10$, $\alpha_1=10^{-3}$, $\alpha_2=10^{-4}$.
  }
\label{local}
\end{figure}

\section{Results}
\subsection{Phases under local predator-prey interactions}
This reaction-diffusion system (Eq.~\ref{asymmetry}) presents a homogeneous fixed point of population densities, $A_s=\lambda_2/(\alpha_2\Omega_2)$ and $B_s=\lambda_1/(\alpha_1\Omega_1)$, where $\Omega_1=\int \mathrm{d}\vec{r}S_1(|\vec{r}|;\Ri)$ and $\Omega_2=\int \mathrm{d}\vec{r}S_2(|\vec{r}|;\Ra)$ are respectively the shape-space volume of the ``inhibition sphere" centered at an antigen and that of the ``activation sphere" surrounding a receptor. When receptor-antigen interactions are local, depending on the ratio of the rates $\lambda_1/\lambda_2$ and diffusivity $D_1/D_2$, coevolving populations exhibit two main phases within the chosen parameter range (Fig.~\ref{local}A): antigen early extinction (colored region) and persistence (white region); the latter divides into two subphases, steady traveling waves (upper) and uniform coexistence (lower); as seen in typical kymographs of the 1D density fields (insets), starting from localized antigens and uniform receptors.

Extinction is expected when antigens replicate fast (large $\lambda_1/\lambda_2$) but mutate slowly (small $D_1/D_2$): after a brief delay during which antigen reaches a sufficient prevalence to trigger receptor proliferation, receptors rapidly expand in number and mutate to neighboring types; the pioneer receptors stay ahead of mutating antigens and eliminate them before escape mutants arise. Once antigen is cleared, the receptor population regresses to the resting level (Fig.~\ref{local}B upper panel). With sufficiently high replication rates, faster mutation allows antigen mutants to lead the arms races against receptors resulting in a persistent evolving state --- a traveling wave Red Queen state, similar to that shown in a recent model of influenza evolution under cross-immunity~\cite{yan:19}. Interestingly, as $\lambda_1/\lambda_2$ increases, while the rate of extinction (color-coded in the phase region) increases, a smaller $D_1/D_2$ is needed for transition to the traveling wave state. On the other hand, at modest $\lambda_1/\lambda_2$,
a uniform coexistence phase is reached following population cycles dampened by mutation (Fig.~\ref{local}B lower panel).
Under local interactions, this homogeneous fixed point is stable to perturbation and does not support spontaneous antigen speciation (i.e., breakup of a continuum into fragments in the shape space). Thus, in what follows, we start from this uniform steady state and introduce the key ingredient --- asymmetric nonlocal interaction --- to show how it drives spontaneous organization.


\begin{figure}[t]
  \includegraphics[width=1\columnwidth]{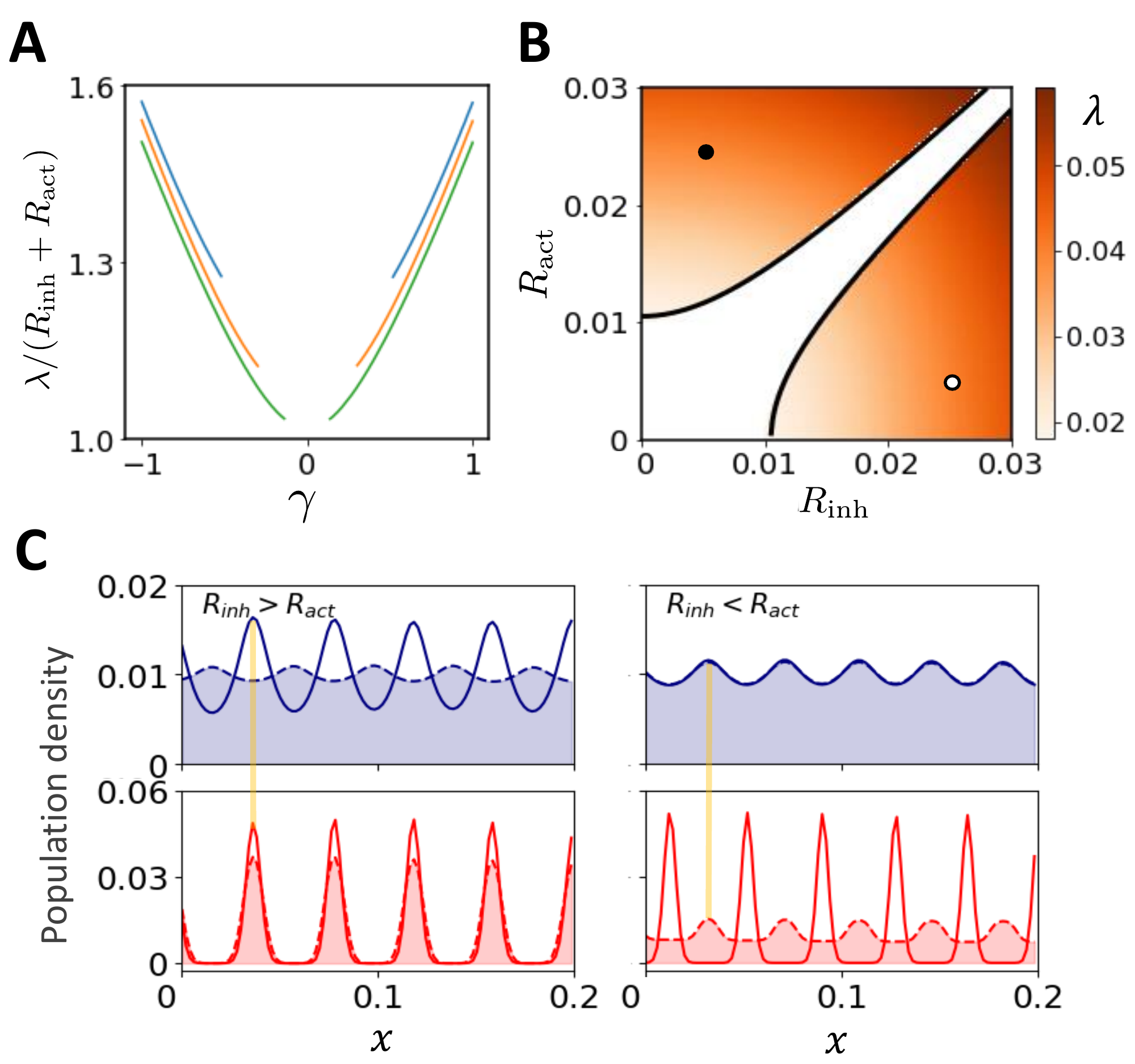}
  \caption{Asymmetric cross-reactive interactions simultaneously organize receptor and antigen distributions. (A and B) The pattern wavelength, $\lambda$, identical for both populations, is symmetric under the interchange of the interaction ranges $\Ri$ and $\Ra$. (A) The scaled wavelength increases with the extent of asymmetry $\gamma\equiv(\Ri-\Ra)/(\Ri+\Ra)$; $\Ri+\Ra=0.015, 0.02, 0.03$ from top to bottom. (B) Pattern diagram in the ($\Ra,\Ri$) plane. The white region corresponds to stable behavior, whereas patterning occurs in the colored areas. Solid lines indicate the instability onset (Eq.~\ref{boundary}). The color bar shows the values of the wavelength determined from the critical mode.
  (C) Typical mutual distributions of receptor (blue) and antigen (red) in a 1D trait space with coordinate $x$. The actual (solid line) and effective (dashed line) population densities (scaled by total abundance) show mismatch for receptors (antigens) when $\Ri>\Ra$ ($\Ri<\Ra$), leading to colocalized (alternate) density peaks between two populations, as indicated by the yellow bars. Shaded are the effective density fields $A_{\mathrm{eff}}(x)$ and $B_{\mathrm{eff}}(x)$. These two examples correspond to the open circle ($\Ri=0.025$, $\Ra=0.005$) and the filled circle ($\Ri=0.005$, $\Ra=0.025$) in panel B.
  $\lambda_1=10$, $\lambda_2=1$, $\alpha_1=10^{-3}$, $\alpha_2=10^{-4}$, $D_1=10^{-6}$, $D_2=10^{-4}$.
  }
\label{pattern}
\end{figure}

\subsection{Simultaneous patterning under asymmetric cross-reactivity}
The analogy between antigen-immunity interaction and predation has been made before~\cite{fenton:10, kimberly:14, armita:16, bradde:17}; however, spontaneous speciation has not been described yet.
On the other hand, for general activator-inhibitor systems --- in the physical space --- Turing patterns can emerge, either from demographic stochasticity which prevents the system from reaching its homogeneous fixed point~\cite{rogers:12, karig:18}, or, more classically, from prohibitively large differences in diffusivity between the autocatalytic and the inhibitory reactants~\cite{turing:52}. Here, using a simple phenomenological model accounting for cross-reactivity (Eq.~\ref{asymmetry}), we show that coevolutionary speciation is possible, without requiring any of the aforementioned patterning mechanisms.

\begin{figure*}[htb]
  \includegraphics[width=1.8\columnwidth]{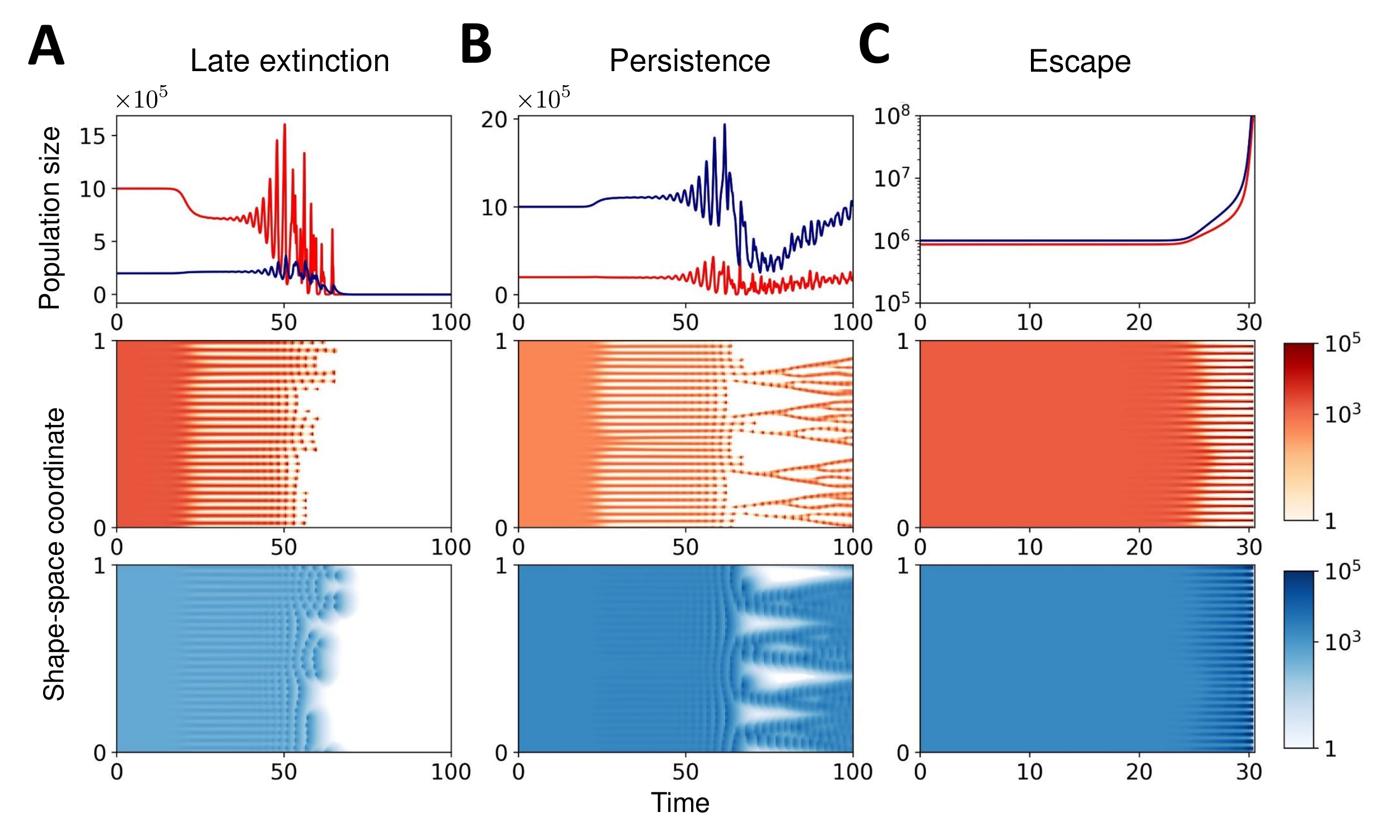}
  \caption{Distinct regimes of coevolutionary dynamics. Population trajectories (top row) and concomitant pattern evolution (lower rows) of antigen (red) and receptor (blue) are shown for late antigen extinction (A), persistent coexistence (B) and antigen escape (C), which are realized by varying the range of cross-reactivity and the size of carrying capacity. Concentration changes progress via three distinct stages: uniform steady state, stationary pattern, and oscillatory pattern. An extinction threshold is crucial for the termination of branches (A, B) and the formation of forks (B) shown in the evolutionary kymographs. Color bars code for population densities.
  (A) $\Ri=0.025$, $\Ra=0.005$, $\theta_2=\infty$; (B) $\Ri=0.005$, $\Ra=0.025$, $\theta_2=\infty$; (C) $\Ri=0.005$, $\Ra=0.025$, $\theta_2=3.5\times10^5$.
  Other parameters are identical to those in Fig.~\ref{pattern}.
  }
\label{phases}
\end{figure*}

To identify the onset of patterning instability, we perturb the uniform stationary state $(A_s, B_s)$ with non-uniform variations, $A(\vec{x},t),B(\vec{x},t)\sim\exp\left(\omega_k t+i\vec{k}\cdot\vec{x}\right)$, where $\vec{k}$ is the wavevector of a spatial Fourier mode. The Turing instability occurs when the most unstable mode, i.e., the critical mode with a wavenumber $k_c$, begins to grow while all other modes decay. This condition, $\mathrm{Re}\left[\omega(k_c)\right]\geq0$, corresponds to the following inequality in the n-dimensional recognition space:
\be
\frac{D_1D_2}{\lambda_1\lambda_2}\le-\frac{1}{k_c^4}
\frac{\hat{S}_1(k_c)\hat{S}_2(k_c)}{\hat{S}_1(0)\hat{S}_2(0)},
\label{patterning}
\ee
where $\hat{S}_1(k)$ and $\hat{S}_2(k)$ are the Fourier transform of the interaction kernels.
It immediately follows that Turing instability in our system is purely driven by \emph{asymmetric nonlocal} interactions and independent of diffusion: if the kernels were symmetric, i.e., $S_1(r; \Ri)=S_2(r; \Ra)$, the right hand side of Eq.~\ref{patterning} can never be positive and hence patterns do not develop; on the other hand, when $D_1 D_2=0$, the patterning condition is most readily satisfied, implying that diffusion is not necessary.
In fact, the commonly assumed Gaussian kernel represents a marginal case which does not robustly warrant instability~\cite{pigolotti:07}. Instead, $\hat{S}(k)<0$ is guaranteed if the strength of interaction decreases steeply with increasing separation across the edge of the interaction range [SI]. For simplicity, we assume step-function kernels, $S_1(r)=\Theta(\Ri-r)$ and $S_2(r)=\Theta(\Ra-r)$.
Accordingly, under a modest extent of asymmetry, i.e., $\gamma\equiv(\Ri-\Ra)/(\Ri+\Ra)\ll1$, the pattern-forming condition can be explicitly expressed in terms of $\gamma$ [SI]:
\be
|\gamma|\geq\gamma_c\equiv\frac{C_n}{(\Ri+\Ra)^2}\sqrt{\frac{D_1D_2}{\lambda_1\lambda_2}},
\label{gamma_c}
\ee
or equivalently,
\be
|\Ra^2-\Ri^2|\geq C_n\sqrt{\frac{D_1D_2}{\lambda_1\lambda_2}}.
\label{boundary}
\ee
Here $C_n$ is a constant that only depends on the dimension, $n$, of the shape space. Rapid increase of $C_n$ with $n$ (Fig.~S3) indicates that stable uniform coexistence extends to stronger asymmetry as the phenotypic space involves higher dimensions.
Therefore, under sufficient asymmetry, a continuum of antigen (receptor) types spontaneously segregates into species-rich and species-poor domains with densities on either side of $A_s$ ($B_s$).
The spacing between adjacent antigen or receptor density peaks, i.e., the pattern wavelength $\lambda\simeq 2\pi/k_c$, is modestly larger than the sum of activation and inhibition radii (Fig.~\ref{pattern}A) due mainly to asymmetry and slightly to diffusion. Note that the minimum level of asymmetry required for patterning decreases with increasing range of cross-reactivity as $\gamma_c\sim(\Ri+\Ra)^{-2}$ (Eq.~\ref{gamma_c}); furthermore,
the pattern wavelength is symmetric under the interchange of $\Ra$ and $\Ri$ (Figs.~\ref{pattern}A and \ref{pattern}B).




However, mutual distributions of receptor and antigen break the symmetry (Fig.~\ref{pattern}C): co-localized patterns form when $R_{\mathrm{act}}<R_{\mathrm{inh}}$ (left panel) while alternate patterns emerge when $R_{\mathrm{act}}>R_{\mathrm{inh}}$ (right panel).
This seemingly counterintuitive behavior can be explained by a rather general mechanism.
When $2\Ri>\lambda$, the ``inhibition sphere" of an antigen may enclose adjacent receptor density peaks. As a result, locations between the peaks, where the actual receptor density $B(x)$ (blue solid line) is in fact the lowest, are instead the worst positions for antigens to be in, because the effective receptor density field acting on antigens at position $x$, $B_{\mathrm{eff}}(x)=\int_{x-\Ri}^{x+\Ri}B(y)\mathrm{d}y$ (blue dashed line), is maximal when $x$ is right amid receptor peaks. Thus the antigen distribution winds up tracking the receptor distribution (yellow bar, left panel). Conversely, when $2\Ra>\lambda$, the ``activation sphere" of a receptor may encompass adjacent antigen peaks; the stimulation for receptor replication is strongest in between the peaks, according to the effective antigen densities $A_{\mathrm{eff}}(x)=\int_{x-\Ra}^{x+\Ra}A(y)\mathrm{d}y$ (red dashed line). Consequently, receptors view antigens as most concentrated in positions where they are actually least prevalent (red solid line).
Therefore, depending on whether $\Ri$ or $\Ra$ is larger, colocalized or alternate distributions result, which reflect a mismatch between the actual distribution and the effective one seen by the apposing population.
In what follows we show that distinct spatial phase relations between mutual distributions will lead to drastically different pattern dynamics and evolutionary outcomes.

\begin{figure*}[t]
  \includegraphics[width=2\columnwidth]{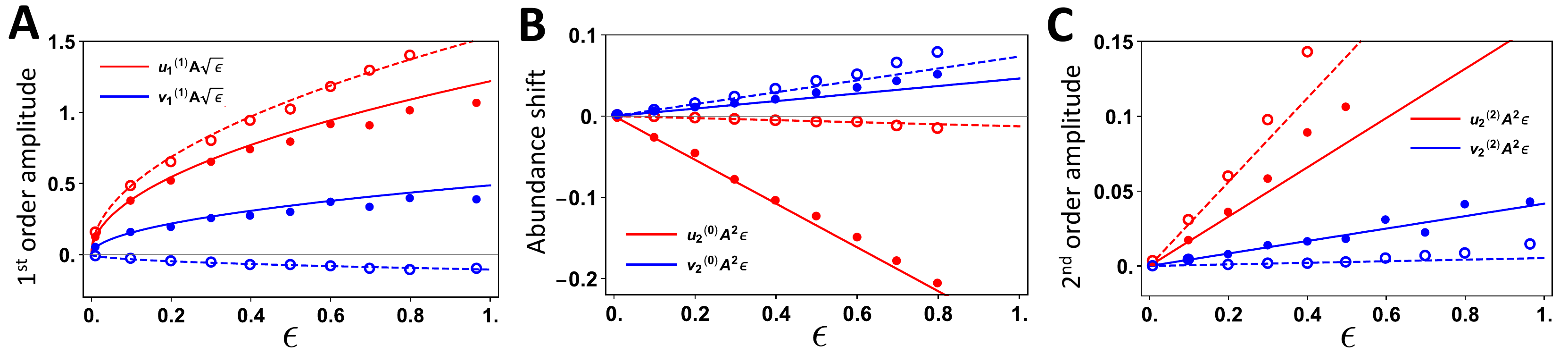}
  \caption{Theory predicts pattern amplitudes and abundance shift induced by coupling between Turing modes. Shown are scaled first (A) and second (C) order pattern amplitudes and abundance shift (B) as a function of $\epsilon$, the dimensionless deviation from $D_1^*$. Lines are analytical predictions; symbols are numerical solutions.
  Solid line and filled symbol: $\Ri=0.025$, $\Ra=0.005$; dashed line and open symbol: $\Ri=0.005$, $\Ra=0.025$. Red (blue) for antigen (receptor). $\theta_2=\infty$.
  }
\label{nonlinear}
\end{figure*}

\subsection{Coevolutionary regimes and ecological feedback}

\begin{figure}[t]
  \includegraphics[width=0.69\columnwidth]{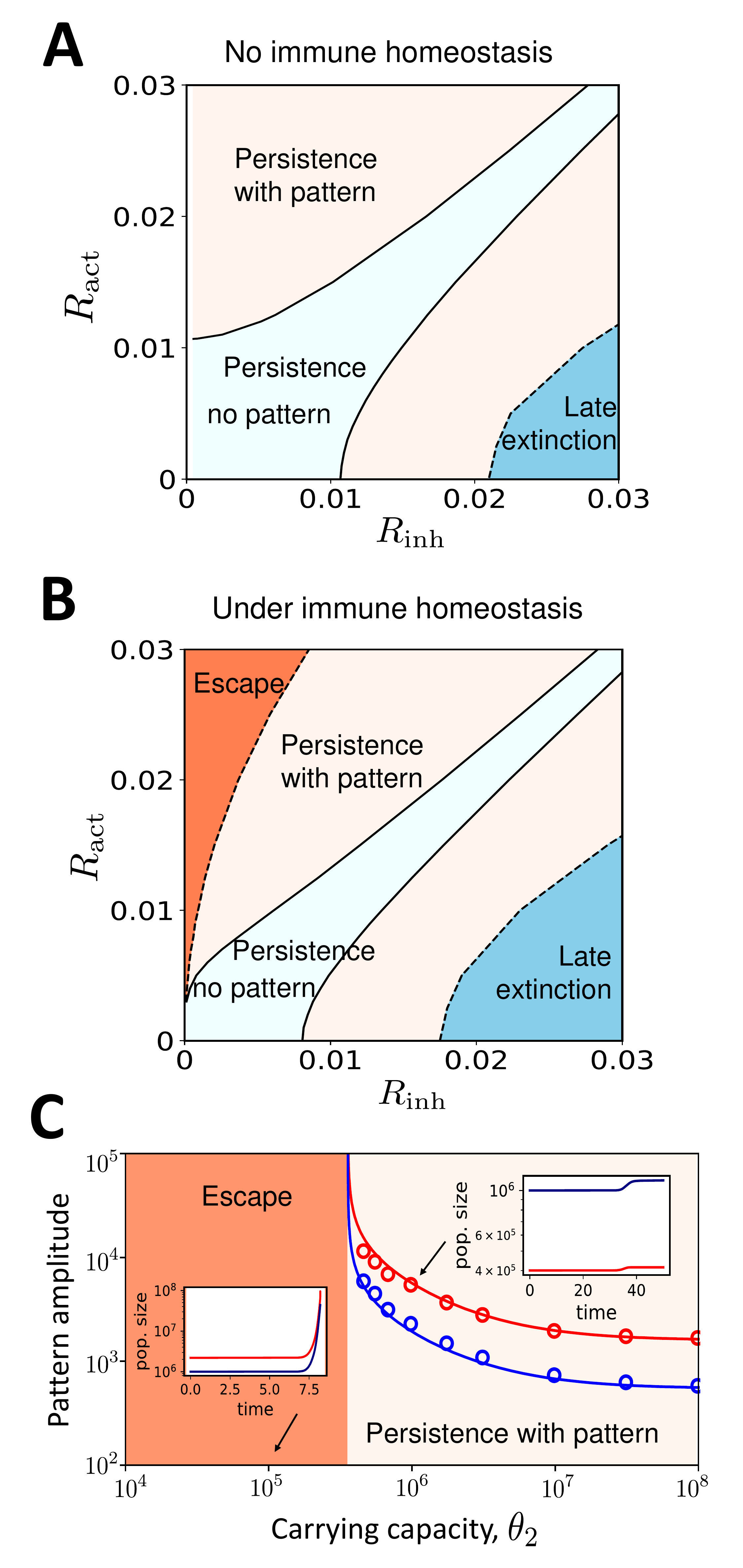}
  \caption{Asymmetric cross-reactivity yields diverse phases.
  (A) Without homeostatic constraints on lymphocyte counts ($\theta_2=\infty$), above the critical asymmetry (beyond the light blue region), patterns form. The pattern-forming boundaries are symmetric about the diagonal. The boundary between the late antigen extinction phase (blue) and the persistent patterned phase (yellow) is determined by tracking the prevalence trajectories until $t=100$. (B) Under a finite carrying capacity ($\theta_2=3\times10^5$), the pattern-forming region is no longer symmetric and an antigen escape phase (red) emerges at the small-$\Ri$ large-$\Ra$ corner, where the phase boundary corresponds to the transition between supercritical and subcritical bifurcations. (C) First order pattern amplitudes as a function of carrying capacity $\theta_2$. Lines are analytical solutions of amplitude equations, and symbols are numerical values extracted from Fourier spectrum of stationary patterns right after abundance shift. Insets show examples of population dynamics in escape (subcritical) and persistence with pattern (supercritical) phases; pattern amplitudes diverge near the transition. Red (blue) for antigen (receptor).
  }
\label{diagram}
\end{figure}

\noindent\textbf{Multi-stage patterning.}
Shown in Fig.~\ref{phases} are the abundance trajectories (top row) and kymographs of concurring patterns (lower rows) demonstrating their concomitant progression and mutual influence. Depending on the sign of asymmetry and the size of carrying capacity, qualitatively distinct regimes appear, including late antigen extinction, persistence, and escape (panels A to C). Intriguingly, concentrations and patterns evolve via three distinct stages: uniform steady state, stationary pattern, and oscillatory pattern. Right after co-patterns spontaneously emerge from the homogeneous steady state, concentration changes in both populations are observed: in the absence of homeostatic constraints (panels A and B), antigen abundance (red) shifts downward whereas receptor prevalence (blue) shifts upward. This is unanticipated because patterning instability in a density field is not expected to alter the overall abundance: growing unstable modes merely redistribute densities in space without changing the average concentration. This appears to break down when patterns develop in two interacting density fields. In fact, the most unstable modes (with wavenumber $k_c$) from both populations couple and modify the zero modes, resulting in the shift in mean population densities.
A weakly nonlinear analysis close to the critical point quantitatively captures both the phase relation between patterned distributions and the shift in overall abundances (Fig.~\ref{nonlinear}). For analytical tractability, we perform the calculation in 1D [see SI for details]. Below we only stress the essential results.

\noindent\textbf{Co-localized and alternate quasispecies.}
Close to the patterning transition, $D_1=D_1^\star(1-\epsilon)$, where $D_1^\star$ is the critical diffusion constant of antigen and $\epsilon$ is small and positive, we seek stationary solutions of the form $A(x)=A_s(1+u(x))$ and $B(x)=B_s(1+v(x))$, where the deviation $\bm{w}=(u(x), v(x))^{T}$ from the homogeneous steady state $(A_s, B_s)^{T}$ is expanded in powers of $\epsilon^{1/2}$ to the second order:
\begin{equation}
\bm{w}=\bm{w}_1^{(1)}\cos(k_cx)\mathcal{A}\epsilon^{1/2}+\left(\bm{w}_2^{(0)}
+\bm{w}_2^{(2)}\cos(2k_cx)\right)\mathcal{A}^2\epsilon.
\end{equation}
The saturated amplitude $\mathcal{A}$ of the perturbation is determined by the amplitude equation at the order of $\epsilon^{3/2}$. The spatial phase difference between the leading pattern modes in antigen and receptor populations, $u_1^{(1)}$ and $v_1^{(1)}$, respectively, can be found from
\begin{equation}
\xi\equiv\frac{v_1^{(1)}}{u_1^{(1)}}=\frac{\lambda_2}{D_2k_c^2}\frac{\sin(k_c\Ra)}{k_c\Ra}
=-\frac{D_1^\star k_c^2}{\lambda_1}\frac{k_c\Ri}{\sin(k_c\Ri)},
\end{equation}
which implies that $\xi\propto\sin\left(\frac{2\pi}{\lambda}\frac{(\Ra+\Ri)}{2}(1-\gamma)\right)
=\sin\left(\pi\frac{1-\gamma}{\lambda/(\Ra+\Ri)}\right)$, with $\lambda$ being the pattern wavelength and $\gamma=(\Ri-\Ra)/(\Ri+\Ra)$. It immediately follows that when $\gamma<0$ (i.e., $\Ra>\Ri$), $1-\gamma>\lambda/(\Ra+\Ri)>1$ (see Fig.~\ref{pattern}A), thus $\xi<0$; when $\gamma>0$ (i.e., $\Ri>\Ra$), $1-\gamma<1<\lambda/(\Ra+\Ri)$, thus $\xi>0$. Therefore, the spatial patterns of antigen and receptor distributions are either in phase ($\xi>0$) or out of phase ($\xi<0$), purely determined by the sign of asymmetry (Fig.~\ref{nonlinear}A). This provides rigor to the intuitive argument we made earlier in relation to Fig.~\ref{pattern}C.
Furthermore, the changes in the overall abundance of antigens and receptors are proportional to $u_2^{(0)}$ and $v_2^{(0)}$, respectively. At $\mathcal{O}(\epsilon)$, we find $u_2^{(0)}\propto-\xi\sin(k_c\Ra)<0$ and $v_2^{(0)}\propto-\xi\sin(k_c\Ri)>0$, i.e., the direction of abundance shift is independent of the sign of asymmetry, in line with numerical solutions (Figs.~\ref{phases}A and \ref{phases}B, top row; Fig.~\ref{nonlinear}B). Importantly, $u_2^{(0)},v_2^{(0)}\propto u_1^{(1)}v_1^{(1)}$, indicating that shift in abundance indeed results from coupling between simultaneous Turing modes.



\noindent\textbf{Dynamic transients.}
A further surprise comes at longer times: the stationary co-patterns are only metastable. Soon after abundance shift takes place, instability starts to grow, visible as increasingly strong oscillations that eventually drive the antigen population to pass below the extinction threshold (Fig.~\ref{phases}A top panel). By perturbing around the abundance-shifted stationary patterns, we indeed identify a growing oscillatory instability from the dispersion relation of linearized dynamics [SI].
The interrupted oscillation amplitudes at later times arise from asynchronous extinction of local antigen clusters (Fig.~\ref{phases}A middle panel). Note that this late extinction phase only occurs to colocalized population densities, i.e., when $\Ri>\Ra$.

Upon interchange of $\Ri$ and $\Ra$ (Fig.~\ref{phases}B), pattern evolution exhibits new features: as some antigen clusters go extinct as a result of oscillatory instability, neighboring clusters migrate to these just vacated sites, where receptors decay due to a lack of stimulation and delay in response, thus locally and temporarily evading immune inhibition.
These surviving clusters then go through successive branching events (i.e., widening then splitting), forming a tree-like structure over time. The coevolving receptor population drives the branching and subsequently traces the newly formed branches (see Movie a for pattern dynamics).
Such coevolutionary speciation, enabled by mutation, persists for extended periods of time so that it effectively overcomes the growing oscillations and maintains antigen at modest prevalence indefinitely.
Note that the persistent ramifying pattern only emerges from alternate density peaks, i.e., when $\Ra>\Ri$.

Is there a chance that antigen population can achieve a global escape from immune control as is often envisaged as a catastrophic failure? This does happen as soon as we turn on a sufficiently strong homeostatic constraint on receptor abundance (Fig.~\ref{phases}C).
Considering homeostasis, the decay rate of receptors becomes $\lambda_2 B(x,t)\left(1+\int_{0}^{1} B(y,t)\mathrm{d}y/\theta_2\right)$, which now includes a contribution from the global constraint characterized by a carrying capacity $\theta_2$. Strikingly, reducing the immune capacity appears to alter the nature of the instability (Fig.~\ref{diagram}C): a critical value of $\theta_2$ marks the transition from supercritical bifurcation (yellow region), where nonlinearity acts to saturate the amplitude of the perturbation, to subcritical bifurcation (red region) where higher order processes have to intervene for stabilization. The latter corresponds to the antigen escape phase (Fig.~\ref{phases}C): an unrestrained growth indicates a loss of immune control.

\noindent\textbf{Higher dimension.}
Similar progression of patterns and population dynamics in distinct regimes is also seen in 2D starting from the uniform steady state (Movies b to d). An analogous ``branching" scenario in the persistence phase is particularly intriguing: antigen droplets deform and migrate to neighboring vacant loci and resist elimination. Oscillations of dense spots in both populations resemble the ``twinkling eyes" pattern proposed for synthetic materials. It has been suggested~\cite{yang:03} that oscillatory patterns can arise in a system consisting of two coupled reaction-diffusion layers, one capable of producing Turing patterns while the other supporting Hopf instability. Distinct from these built-in mechanisms, instabilities in our system are self-generated: interacting populations spontaneously fragment in the trait space, and spatial resonance of the resulting Turing modes leads to abundance shift and subsequent growing oscillations.

\noindent\textbf{Phase diagram.}
To stress the role of asymmetric cross-reactivity in governing the diverse behaviors, we present phase diagrams on the $(\Ra, \Ri)$ plane (Fig.~\ref{diagram}). Without homeostatic constraints ($\theta_2=\infty$, panel A), patterns form above the critical asymmetry marked by solid lines that are symmetric about the diagonal; the enclosed patternless phase (light blue region) corresponds to stable homogeneous coexistence like for local interactions. On the $\Ri>\Ra$ side, the late extinction phase (blue) transitions to the persistent patterned phase (yellow) at a boundary (dashed line) determined by tracking the prevalence trajectories until $t=100$; longer tracking time would expand the extinction phase. Under a finite carrying capacity ($\theta_2=3\times10^5$, panel B), two major changes occur: First, the pattern-forming region is no longer symmetric but expands on the $\Ra>\Ri$ side toward the diagonal. Second, the antigen escape phase (red region) emerges at the small-$\Ri$ large-$\Ra$ corner; the phase boundary corresponds to the transition between supercritical and subcritical bifurcations in the amplitude equation. The escape phase enlarges as the carrying capacity diminishes. Thus, our model predicts expansion of the antigen escape phase with age, owing to diminishing counts of renewable lymphocytes~\cite{miller:96}.
The regime of persistence with pattern (yellow), irrespective of the homeostatic constraint, differs between the flanks; while oscillations occur on both sides off the diagonal, antigen branching (Fig.~\ref{phases}B) only appears when $\Ra>\Ri$, manifesting the potential for evasion.



\section{Discussion}
Environment becomes a relative concept in the case of coevolution.
We present a general model of mutual organization between continuous distributions of antigens and receptors that interact cross-reactively. In a shared phenotypic space, the receptor repertoire and antigen population constitute each other's environment and adapt to mutually constructed fitness seascapes. This phenomenological approach allows us to describe the interplay between ecological and evolutionary processes that do not separate in timescales, thus revealing a variety of long-lived transients and dynamic steady states observed in nature, such as antigen extinction, chronic persistence, and unrestrained growth until eventual saturation.

This simple model of antigen-immunity coevolution demonstrates a new type of Turing mechanism that stems from tolerance for imperfect lock-key interactions and disparate conditions for receptor activation and antigen removal. While it might be intuitive that under reciprocal cross-reactivity, antigen and receptor populations would simultaneously fragment (Fig.~\ref{pattern}), more surprises come after the co-pattern emerges (Fig.~\ref{phases}): When two density distributions are in phase ($\Ra<\Ri$), spatial resonance between the lowest Turing modes precedes growing oscillations in the overall abundance, driving antigens to extinction; when apposing populations are out of phase ($\Ra>\Ri$), strong homeostatic constraints on immune cells alter the nature of pattern instability from supercritical to subcritical, leading to uncontrolled growth. The intuitive picture is, when $\Ra<\Ri$, antigens are inhibited by receptors that they do not activate and hence fail to evade immune attack; when $\Ra>\Ri$, receptors are activated by antigens that they cannot inhibit, thus, under resource limits, an increasingly weaker defense results.
Such multi-stage patterning and its feedback to population dynamics, triggered by asymmetric non-local interactions, is a qualitatively new phenomenon, and is clearly distinct from speciation due to competitive exclusion in a single population.
Our predictions are supported by experiments: strong oscillations in antigen abundance prior to crash to extinction have been seen in viral evolution within humans and attributed to cross-reactive antibody response~\cite{bailey:17}, whereas strategies of distracting immune attention are indeed used by many viruses that create a vast excess of defective particles than functional ones~\cite{huang:70}.

Stochasticity arising from demographic noise does not modify qualitative model behaviors in all distinct regimes [SI]. Albeit not required for pattern formation, stochastic fluctuations appear to accelerate the growth of instability and speed up antigen extinction (Fig.~S12); this observation and further effect of demographic noise will receive a careful examination in future work.

Our results suggest that, the immune system may have evolved to exploit the asymmetry between activation and inhibition by differentiating these processes physically and biochemically.
A remarkable example is affinity maturation of B lymphocytes~\cite{victora:12} in which rapid Darwinian evolution acts over days to weeks to select for high affinity clones: Immature B cells are trained in lymphoid tissues where antigens are presented in a membrane form and declining in availability; fierce competition for limited stimuli thus provides a sustained selection pressure that constantly raises the activation threshold, i.e., decreasing $\Ra$. In contrast, mature B cells then released into circulation encounter soluble antigens at higher abundance, corresponding to $\Ri>\Ra$. As a result, enhanced asymmetry between conditions for immune stimulation and antigen removal facilitates elimination of pathogens.
Conversely, pathogens evolve immunodominance~\cite{schrier:88} and make fitness-restoring mutations~\cite{crawford:07} that increase $\Ra$ and decrease $\Ri$, both of which aid in evasion.


A dynamic balance hence persistent coevolution, often pictured as an asymptotic state, can only be sustained when the asymmetry is not too severe. It is thus likely to be favorable if the extent of asymmetry stays near the edge between persistence and imbalance, which adjusts to the tension between the need for defense against foreign pathogens ($\gamma>\gamma_c$) and that for tolerance toward benign self tissues ($|\gamma|\leq\gamma_c$).
Interestingly, the critical asymmetry $\gamma_c$ rises with the dimension of the phenotype space (Fig.~S3), suggesting that dynamic balance become easier to maintain for more complex traits.
Our analysis also reveals other stabilizing factors for coexistence [SI], including a stronger influx of naive cells, larger jump sizes in trait values, and longer-tailed interaction kernels.


Because the present model of antigen-immunity coevolution is a sufficiently abstracted one,
having properties which seem quite robust and independent of the details of predation, we expect that the results and predictions are relevant for a wide range of coevolving systems including cancer cells and T lymphocytes, embryonic tissues and self-reactive immune cells, as well as bacteria and bacteriophage. This model can be adapted to be more biologically faithful, e.g., by incorporating preexisting antigenic landscapes, taking rates to be age-dependent, and treating cross-reactivity as an evolvable character.

Our predictions can potentially be tested in experiments that track both the pathogen load and the diversity history via high-throughput longitudinal sequencing of receptors and antigens~\cite{liao:13, gao:14, moore:15, freund:17}. In addition, phenotypic measurements using binding and neutralization assays~\cite{west:13} can inform the extent of asymmetry. Combining these two sets of experiments in different individuals would allow to correlate the degree of asymmetry with evolutionary outcomes.




We hope that this work proves useful in providing a framework for understanding and testing how cross-reactive interactions --- ubiquitous and crucial for biological sensory systems --- can lead, in part, to the generation, maintenance and turnover of diversity in coevolving systems. More broadly, our work provides the basis for a theory of evolution in responsively changing environments, highlighting that ecological feedback in pattern-forming systems can yield dynamic transients and drive evolution toward non-steady states that differ from the Red Queen persistent cycles.

\textit{Acknowledgments}.---We thank Sidney Redner and Paul Bressloff for enlightening discussions. SW gratefully acknowledges funding from UCLA.


\section{Appendix}

Below we provide additional description of the main steps in our analysis. The three subsections are ordered in accordance with the successive instabilities that give rise to the multi-stage progression of trait-space patterns and associated population dynamics.

\subsection{Linear instability of the uniform steady state}

To identify the onset of patterning instability, we linearize the equation of motion (Eq.~1 in the main text) around the homogeneous fixed point $(A_s, B_s)$ and work in the Fourier space.
Defining $A(\bm x, t)=A_s+\sum_k\delta A_k\exp(\omega_k t+i\bm k\cdot \bm x)$ and $B(\bm x, t)=B_s+\sum_k\delta B_k\exp(\omega_k t+i\bm k\cdot \bm x)$ yields
$$
\omega(k)
\begin{pmatrix}
\delta A_k\\
\delta B_k\\
\end{pmatrix}
=
\begin{pmatrix}
-D_1 k^2 & -\alpha_1 A_s \hat{S}_1(k)\\
\alpha_2 B_s \hat{S}_2(k) & -D_2k^2\\
\end{pmatrix}
\begin{pmatrix}
\delta A_k\\
\delta B_k\\
\end{pmatrix},
$$
where $\hat S_1(k)$ and $\hat S_2(k)$ are the Fourier transform of interaction kernels $S_1(r)$ and $S_2(r)$, respectively. Solving this characteristic equation gives the dispersion relation, i.e., the linear growth rate of the Fourier modes:
$$
 \omega(k)=-(D_1+D_2)k^2+\sqrt{(D_1-D_2)^2k^4
 -4\lambda_1\lambda_2\frac{\hat S_1(k)\hat S_2(k)}{\hat{S}_1(0)\hat{S}_2(0)}}.
$$
Turing instability occurs when the least stable mode (with a wavevector $k_c$) begins to grow, namely,
$$
Re[\omega(k_c)]\geq0,
$$
where the wavenumber of the critical mode can be determined by $\partial_k \omega|_{k=k_c}=0$. This gives the pattern-forming condition (Eq.~\ref{patterning} in the main text).

\subsection{Weakly nonlinear stability analysis: amplitude of patterns}

In this section we lay out the procedure of amplitude expansion~\cite{gambino:13, stephenson:95} that yields an approximate description of the patterns formed.

Close to the transition, we introduce a small positive parameter $\epsilon=(D_1^\star-D_1)/D_1^\star$. Thus, terms in the equation of motion can be collected into a linear part evaluated at the transition, $\mathcal{L}^\star\bm w$, and the rest, $\bm g_{\epsilon}(\bm w)$, for which the evolution operator explicitly depends on $\epsilon$:
$$
\partial_t\bm w=\mathcal{L}^\star\bm w+\bm g_{\epsilon}(\bm w).
$$
Here $\bm w=(u(x), v(x))^T$ denotes the deviation from the homogeneous steady state.

In the vicinity of the transition, one can expand the inhomogeneous deviation in powers of $\epsilon^{1/2}$:
$$
\bm w=\bm w_1\epsilon^{1/2}+\bm w_2\epsilon +\bm w_3\epsilon^{3/2} +o(\epsilon^2).
$$
Note $\bm w_n\propto\mathcal{A}^n$, where $\mathcal{A}$ represents pattern amplitude; the closer to the onset of instability, the smaller the saturated amplitude of the perturbation.
Substituting this expansion into the equation of motion yields order by order (explicit expressions can be found in SI):
$$
\begin{aligned}
&O(\epsilon^{1/2})
&&\quad\quad
\mathcal{L^\star}\pmb{w}_1=0\\
&O(\epsilon)
&& \quad\quad
\mathcal{L^\star}\pmb{w}_2=\pmb{F}(\pmb{w}_1)\\
&O(\epsilon^{3/2})
&&\quad\quad
\mathcal{L^\star}\pmb{w}_3=\pmb{G}(\pmb{w}_1, \pmb{w}_2)
\end{aligned}
$$
The first equation effectively recovers the linear theory.
For nontrivial solutions to these equations to exist, certain conditions need to be met. Such solvability condition associated with the third equation serves to determine
the pattern amplitude $\mathcal{A}$. It reads
$$
\langle\bm\rho^{\dagger}, \bm G\rangle = \int\bm\rho^{\dagger}(x)^T\bm G(x)dx=0,
$$
where $\bm\rho^{\dagger}$ is the nontrivial kernel of the adjoint operator $\mathcal L^{\star\dagger}$, satisfying $\mathcal L^{\star\dagger}\bm\rho^\dagger=0$. This orthogonal condition results in the stationary amplitude equation (or Stuart-Landau equation)
$$
a_0\mathcal A-a_1\mathcal A^3=0,
$$
where $a_0$ is the linear growth rate and $a_1$ the Landau parameter, both of which depend on system parameters.
The formation of steady patterns with a finite amplitude requires $a_0\times a_1>0$, corresponding to supercritical bifurcation where nonlinearity acts to saturate the growth of linearly unstable modes. In contrast, $a_0\times a_1<0$ leads to subcritical bifurcation, indicating that even higher order nonlinearities are required to stabilize the pattern.

This method also applies when homeostasis is considered. Transition from supercritial to subcritical bifurcation occurs when the carrying capacity of immune receptors (predators) falls below a critical value (Fig.~\ref{diagram}C, Fig.~S9).

As such, we obtain an approximate solution of the patterned state close to transition (Eq.~5 in the main text), which shows nice agreement with the numerical solution of the equation of motion (Fig.~\ref{nonlinear}).

\subsection{Pattern stability}

To determine the stability of the patterned state, we introduce a small perturbation $\delta \bm w$ to the steady pattern $\bm w_s$ obtained in the last section, i.e., $\bm w=\bm w_s+\delta \bm w$, so that the perturbation evolves according to
$$
\partial_t\delta \bm w=\mathcal L_s\delta\bm w,
$$
where $\mathcal L_s$ stands for the linearized evolution operator evaluated at $\bm w_s$. Again we expand the perturbation in Fourier modes
$
\delta \bm w=\int_{-\infty}^{\infty}\delta\bm w_q e^{iqx}\textrm{d}q.
$
Since the operation of $\mathcal L_s$ contains modes $k_c$ and $2k_c$ (according to Eq.~5), it couples $q$-mode with those having wavenumbers $q\pm k_c$, $q\pm2k_c$, etc. Consequently, unlike perturbations to the homogeneous state that lead to decoupled characteristic equations for individual modes, now modes that differ by multiples of $k_c$ are all coupled and evolve together according to a matrix equation
$$
\partial_t\bm\Psi=M\bm\Psi.
$$
In principle, the state vector $\bm\Psi$ would be infinite dimensional, containing modes of perturbations associated with wavenumbers $q, q\pm k_c, q\pm2k_c, \cdots$. Correspondingly, the evolution matrix $M$ would also be infinite, forbidden from being solved. Thus, truncation of the cascade is needed to make progress. After truncating the matrix to a modest rank, we can proceed by numerically solving for the eigenvalues of the characteristic matrix.
Then, like in usual stability analysis, the leading eigenvalue, $m_1(q)$, describes the long-term growth of the perturbation. Close to the transition, $m_1(q)$ solved as such agrees well with the linear growth rate of perturbations observed in numerical solutions of the original equation of motion (Fig.~S7).
In particular, this analysis captures the growing oscillations in the patterned state (as shown in Fig.~\ref{phases} in the main text); soon after patterns form in both populations, an oscillatory instability ensues; this linear instability stems from coupling between the spatial modes of the perturbation and those of the steady pattern with matched wavelengths.

%


\end{document}